\documentclass{article}

\usepackage{arxiv}

\usepackage[utf8]{inputenc} 
\usepackage[T1]{fontenc}    
\usepackage{hyperref}       
\usepackage{url}            
\usepackage{booktabs}       
\usepackage{amsmath}
\usepackage{amsfonts}       
\usepackage{nicefrac}       
\usepackage{microtype}      
\usepackage{lipsum}		
\usepackage{graphicx}
\usepackage{natbib}
\usepackage{doi}
\def\PR{\text{PR}}

\def\prob{\mathbb{P}}
\def\dens{f}
\newcommand\denscond[2]{\dens(#1|#2)}
\newcommand\gauss[3]{f_{\mathcal{N}}\left(#1\,;#2,#3\right)}
\def\mix{\text{p}}
\def\setsymbol#1{#1} 

\title{Reconciling risk-based and storyline attribution with Bayes theorem}

\date{July 2024}	

\author{ \href{https://orcid.org/0000-0003-4750-361X}{\includegraphics[scale=0.06]{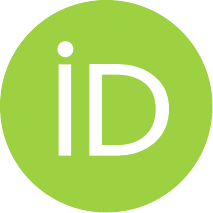}\hspace{1mm}Sebastian Buschow} \\
	Institute of Geosciences, 
	Meteorology Section \\
	University of Bonn \\
	\texttt{sebastian.buschow@uni-bonn.de} \\
	\And
	\href{https://orcid.org/0000-0003-4566-572X}{\includegraphics[scale=0.06]{orcid.pdf}\hspace{1mm}Petra Friederichs} \\
	Institute of Geosciences, 
	Meteorology Section \\
	University of Bonn \\
    \And
     Andreas Hense \\
	Institute of Geosciences, 
	Meteorology Section \\
	University of Bonn \\
}

\renewcommand{\headeright}{}
\renewcommand{\undertitle}{}
\renewcommand{\shorttitle}{Reconciling risk-based and storyline attribution}

\hypersetup{
pdftitle={Reconciling risk-based and storyline attribution with Bayes theorem},
pdfauthor={S. Buschow, P. Friederichs, A. Hense}
}

\begin{document}

\maketitle

\begin{abstract}
    The question to what extent climate change is responsible for extreme weather events has been at the forefront of public and scholarly discussion for years. Proponents of the ``risk-based'' approach to attribution attempt to give an unconditional answer based on the probability of some class of events in a world with and without human influences. As an alternative, so-called ``storyline'' studies investigate the impact of a warmer world on a single, specific weather event. This can be seen as a conditional attribution statement. 
    In this study, we connect conditional to unconditional attribution using Bayes theorem: in essence, the conditional statement is composed of two unconditional statements, one based on all available data (event and conditions) and one based on the conditions alone. 
    We explore the effects of the conditioning in a simple statistical toy model and a real-world attribution of European summer temperatures conditional on blocking. The resulting attribution statement is generally strengthened if the conditions are not affected by climate change. Conversely, if part of the trend is contained in the conditions, a weaker attribution statement may result. 
    
\end{abstract}

\section{Introduction}

More than three decades since the first IPCC report in 1990, improvements in physical understanding, statistical methods and, most simply, the continuation of the observational record, have left practically no room for doubt that observed global warming can in large part be attributed to human activities \citep{lee2023climate}. In view of already observed catastrophic climate extremes, the attention of many attribution studies has shifted to regional scales and individual extreme events. 

Research on so-called event attribution has broadly split into two camps: the original main stream descends from the ideas of \cite{allen2003liability,stott2004human} and compares the probabilities for some event of interest in a world with and without human influence. The techniques and procedures used in this so-called ``risk-based'' approach have reached a degree of standardisation and maturity which allows projects like World Weather Attribution to conduct attribution studies in near real time \citep{philip2020protocol}. Policymakers and the interested public can thus potentially be informed about the human contribution to an extreme event shortly after it makes global headlines. 

Starting with \cite{trenberth2011}, risk-based event attribution has been criticised for its tendency to accept the null hypothesis ``no human impact'' in situations where the quantity of interest is too noisy and the rare weather situation responsible for the observed event may be poorly represented in climate models. To address this shortcoming, the ``storyline'' approach to attribution quantifies the impact of global warming, conditional on the specific weather in space and time  that leads to the extreme event. 

While the risk-based analyses rely either on observed statistics or free running global climate models, storyline studies are typically realised with targeted regional climate simulations. Here,  initial and or boundary conditions that guarantee the occurrence of the event are modified to represent the same event in a warmer or cooler climate. The fundamental divide between the two approaches, however, lies not in the datasets but in the use of conditional statistics \citep{shepherd2016}: risk-based studies consider the statistics of some variable, conditional only on a scenario (human impact yes or no). Storylines are based on the statistics (not necessarily the full probability distribution) of the event given the scenario and some other quantities that represent the dynamical state of the climate system outside of the event. 

Given this view of the two approaches based on different \textit{conditional} statistics, it seems obvious that they can be linked by techniques from Bayesian statistics. \cite{shepherd2016} symbolically draws the connection in terms of conditional probabilities, but does not apply Bayes theorem. Studies that explicitly compute conditional probabilities for attribution are rare. \cite{chiang2021multivariate} implement this idea in a bivariate setting using copulas but do not discuss the connection to storylines. This aspect is touched on more directly by \cite{yiou2017statistical} who decompose dynamic and thermodynamic drivers of an extreme precipitation events. The latter study is an example of the ``analogue-based'' style of event attribution. While this approach is sometimes viewed as a third class besides risk-based and storyline \citep{stott2016attribution, otto2017attribution}, it can also be seen as a special case of conditional attribution. 

In this study, we present a simple and general link between conditional and unconditional attribution studies. This connection is derived in section \ref{sec:theory} using Bayes theorem. Here, we also highlight the connection to storylines and analogue attribution. In section \ref{sec:toy}, we study the basic effects of conditional attribution using a simple statistical toy model. To apply our idea to a real-world attribution problem, a more powerful Bayesian statistical model is introduced in section \ref{sec:method}. As our example, we study central European summer temperatures conditional on blocking activity. Sections \ref{sec:data} and \ref{sec:res} summarise the CMIP6 and ERA5 data used and the conditional and unconditional attribution results. Some concluding remarks and perspectives are given in section \ref{sec:outro}.

\section{Theory}\label{sec:theory}

\begin{table}
    \centering
    \begin{tabular}{l|l}
         $\prob$ & probability \\ 
         $\dens$ & probability density (pdf) \\
         $\gauss{x}{\mu}{\Sigma}$ & Gaussian pdf with mean $\mu$ and covariance $\Sigma$ \\ 
         $F$ & cumulative distribution function (cdf) \\
         \mix & joint probability mass and density \\
         \PR & Probability ratio
    \end{tabular}
    \caption{Notation used throughout this manuscript. }
    \label{tab:notation}
\end{table}

\subsection{Unconditional attribution}
We represent the observed state of the climate system by a real valued random variable $X$ with realisations $x\in\mathbb{R}^q$ and probability density $\dens(x)$. In general, the elements of $x$ could correspond to a single or to multiple variables at one or more points in space and time, so that $X$ is a multivariate, joint spatio-temporal random process. Attribution studies are interested in the probability of an event 
\begin{align}
    X \in \setsymbol{E}\,, \hspace{5pt} \textit{ with }\setsymbol{E}\subset \mathbb{R}^q
\end{align}
in a world with and without climate change. These two worlds are represented by a factual scenario $S_1$ and a counterfactual scenario $S_0$, respectively. The probability of the event under one of the scenarios is  given by
\begin{align}
    \prob(E|S_i) = \int_\setsymbol{E} \denscond{x}{S_i}\,dx\,,
\end{align}
Where $\denscond{x}{S_i}$ denotes the conditional density of $X$ given scenario $S_i$. 
We would then attribute an observed event $E$ to one of these scenarios according to the probability ratio
\begin{align}
    \PR(E) = \frac{\prob(E|S_1)}{\prob(E|S_0)} \in [0,\infty]\,.
\end{align}
$\PR(E)$ is also known as the risk ratio \cite{vanderweele2015}. 
As a typical example, $X$ might be a uni-variate random variable representing a temperature observation at some location. We could define an event $E$ as the exceedance of some threshold $u$. In the notation above, this corresponds to the set $\setsymbol{E}=[u,\infty]$. A probability ratio of $\PR=2$ would then indicate that the risk of temperatures exceeding the threshold $u$ was doubled due to climate change.  

One criticism of this ``risk-based'' approach is that every such attribution study implicitly starts from the null hypothesis ``no climate change'' ($S_0$), even though the \textit{overall} evidence for $S_1$ is already overwhelming \citep{trenberth2011}. Depending on the choice of variable $X$ and event $E$ considered, many attribution studies would appear to accept $S_0$ and reject climate change.   
As hinted at by \cite{trenberth2011} and \cite{shepherd2019storyline}, the apparent discrepancy is immediately resolved if we consider attribution as a Bayesian decision problem \citep{min2004bayesian}. In this approach, the scenario $S_i$ is treated as a binary random variable with a prior probability $\prob(S_i)$, representing our belief in $S_i$ before considering any new evidence. Once we observe the event $E$, our beliefs are updated according to Bayes theorem
\begin{align}
    \prob(S_i|E) = \frac{\prob(E|S_i)\prob(S_i)}{\prob(E)}\,.\label{eq:bayes}
\end{align}
To address the attribution question, a Bayesian would decide between the two scenarios based on the  posterior odds
\begin{align}
    \frac{\prob(S_1|E)}{\prob(S_0|E)}\begin{cases}
    >1\Rightarrow \hspace{5pt}\text{prefer }S_1\\
    \leq 1\Rightarrow \hspace{5pt}\text{prefer }S_0
    \end{cases}\,,
\end{align}
which are related to the probability ratio $\PR$ via
\begin{align}
     \frac{\prob(S_1|E)}{\prob(S_0|E)} = \frac{\prob(S_1)}{\prob(S_0)} \cdot \PR(E) \,.\label{eq:PRbayes}
\end{align}
The ratio $\PR$ thus receives a new interpretation as the so-called Bayes factor (symbol $B$ in \cite{min2004bayesian}) and the decision between $S_0$ and $S_1$ is no longer based on $\PR$ alone. Instead, $\PR$ measures whether observing $E$ added to (or subtracted from) our belief in one of the two scenarios. Values of $\PR>3$ are typically considered ``substantial'' evidence against $S_0$, $\PR>12$ is considered ``strong'', $\PR>150$ ``decisive'' (see \cite{min2004bayesian} and references therein). 

\subsection{Conditional attribution}\label{sub:cond}
A second main criticism raised against the risk based approach  \citep{trenberth2011, trenberth2015, shepherd2019storyline} concerns the role of natural variability in the attribution of extreme events: the weather situation related to atmospheric hazards such as heat waves or pluvial floods is rare by definition. This implies small probabilities which are difficult to estimate and highly uncertain. The solution favoured by these authors is the so-called ``storyline approach'', which answers the attribution question conditional on the weather situation responsible for the event of interest. 

We can formalise a conditional attribution statement by splitting the climate state up into
\begin{align}
    X=(X_1, X_2)\,,
\end{align}
where $X_1$ contains all elements relevant to the \textit{event} and $X_2$ everything else, i.e., the conditions. Accordingly, the event definition reads $X_1\in\setsymbol{E}$ and a certain state of the weather conditions $C$ is defined as $X_2\in\setsymbol{C}$.
We then study the conditional probability ratio
\begin{align}
    \PR(E|C) = \frac{\prob(E|C,S_1)}{\prob(E|C,S_0)}\,.\label{eq:conditionalPR1}
\end{align}
For example $X_1$ could represent winter precipitation in Europe and $X_2$  the NAO index (as in \cite{yiou2017statistical}). Again, we might define the event $E$ as a threshold exceedance $\setsymbol{E}=[u,\infty]$ and the conditions $C$ as the positive phase of the NAO, i.e., $\setsymbol{C}=[0,\infty]$. A conditional $\PR$ of 2 would then imply that climate change is responsible for doubling the chances of heavy precipitation under NAO$^+$ conditions. 

To evaluate equation \ref{eq:conditionalPR1}, it is convenient to re-arrange it with Bayes theorem:
\begin{align}
    \PR(E|C) = \frac{\prob(S_1|E,C)/\prob(S_0|E,C)}{\prob(S_1|C)/\prob(S_0|C)}\,.
\end{align}
On the right-hand side, we recognise two unconditional probability ratios, which yields the concise result
\begin{align}
    \PR(E|C) = \frac{\PR(E,C)}{\PR(C)}\,.\label{eq:conditionalPR2}
\end{align}
To answer the conditional attribution question, we simply need to repeat the Bayesian decision between $S_1$ and $S_0$ twice: once using the knowledge about the complete state, i.e., $X\in\setsymbol{E}\times\setsymbol{C}$, and once using only the conditions $X_2\in\setsymbol{C}$.

\subsection{Continuous multivariate events}\label{sub:conti}
A third possible criticism of risk-based attribution is that events are almost always defined as a uni-variate threshold exceedance, as in our examples above. In contrast, an observed extreme event is often a unique combination of multiple variables arranged in space and time. Such an individual observation can inspire the choice of a single relevant variable and threshold, but otherwise it plays no central role in a risk-based attribution study \citep{shepherd2016}. We will now see that the Bayesian approach naturally ameliorates these limitations. 

First, let us consider the possibility of defining an event in terms of the single realisation of a real univariate random variable that was actually observed, i.e., $\setsymbol{E}=\{x\}$. 
In general, the probability of observing any individual real number (for example a specific temperature in $K$) is zero and the probability ratio (equation \ref{eq:conditionalPR1}) appears to be undefined ($0/0$).
In the univariate case, we recall that the density $\dens$ is the derivative of the probability $\prob$  and use l'Hôpital's rule to find that $\PR$ simply becomes a ratio of pdfs:
\begin{align}
    \PR(x) = \frac{\denscond{x}{S_1}}{\denscond{x}{S_0}}\,.\label{eq:PRdensity}
\end{align}

The multivariate case is more complicated. On the one hand, it is often not obvious how a single multivariate extreme event should be generalised to a class of events: a compound event may result from some specific combination of variables, not all of which are individually extreme. It may therefore be convenient to avoid thresholds and attribute the single observed state instead. On the other hand, it is not obvious what the counterpart to equation \ref{eq:PRdensity} should be since l'Hôpital's rule generally applies only in one dimension. In the Bayesian world, there is no problem since Bayes theorem (equation \ref{eq:bayes}) works just as well for a mixture of discrete and continuous variables:
\begin{align}
    \prob(S_i|x) = \frac{\denscond{x}{S_i}\prob(S_i)}{\dens(x)}\label{eq:bayes_mixed}
\end{align}
If we have a model for the conditional densities, we can thus apply all of the above results to continuous events $X_1=x_1$ and conditions $X_2=x_2$. For example, instead of asking for the probability ratio of heavy rain given NAO$^+$ , we could compute $\PR$ for $50\,mm$ of rainfall given a NAO index of $+1$. Equation \ref{eq:bayes_mixed} and all that follow can be applied to univariate or multivariate $X_1$ and $X_2$ as needed. 

The difference between attribution of a single value ($X=x$, used in this study) and a threshold exceedance ($X>x$, as is typically done in attribution studies) is discussed in more detail in appendix \ref{app:PRcont}.

\section{Simple example cases}\label{sec:toy}

\begin{figure*}
    \centering
    \includegraphics[width=\textwidth]{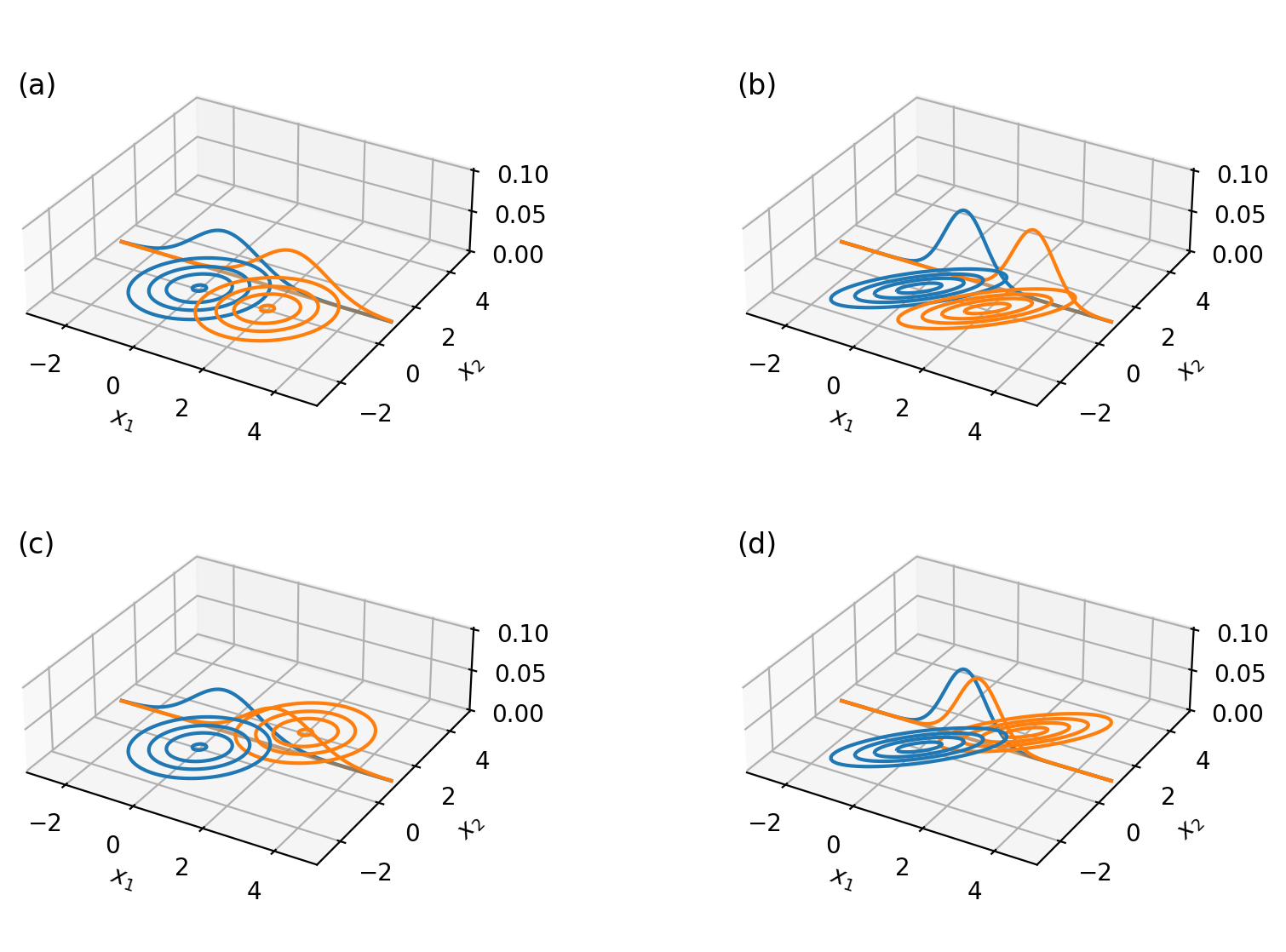}
    \caption{Bivariate normal densities $f(x_1,x_2)$ for $S_0$ (blue) and $S_1$ (orange). Lines above the contours indicate conditional densities $f(x_1|x_2=1)$. In all panels, $\mu_1(S_0)=\mu_2(S_0)=0$ and $\mu_1(S_1)=2$. The other parameters are\\ (a) $\rho=0.2$, $\mu_2(S_1)=0$, (b) $\rho=0.8$, $\mu_2(S_1)=0$, (c) $\rho=0.2$, $\mu_2(S_1)=2$ and (d) $\rho=0.8$, $\mu_2(S_1)=2$.}
    \label{fig:toy}
\end{figure*}

To get an intuition for conditional attribution, let us consider a bivariate state variable $X=(X_1, X_2)$, $x_1,x_2\in\mathbb{R}$ and assume it follows a normal distribution 
\begin{align}
    \gauss{x}{\mu(S_i)}{\Sigma} = \left((2\pi)^2\det\Sigma\right)^{-1/2}\exp\left( -0.5\,(x-\mu(S_i))^T\Sigma^{-1}(x-\mu(S_i))\right)
\end{align}
with expectation value and covariance matrix
\begin{align}
    \mu(S_i) = \begin{pmatrix} \mu_1(S_i) \\ \mu_2(S_i)\end{pmatrix},\hspace{5pt} \Sigma=\begin{pmatrix}
            1 & \rho\\  \rho & 1
        \end{pmatrix}\,.
\end{align}
Here, we have set the variances of both components under both scenarios to unity, in order to focus our analysis on the effect of the correlation $\rho$ between them. Only the expectation values $\mu$ depend on the scenario $S_i$. Figure \ref{fig:toy} illustrates the two bivariate densities for different choices of $\rho$ and $\mu_2(S_1)$.

In this simple example, we can compute the probability ratio for the event $X_1=x_1$ conditional on $X_2=x_2$ directly via equation \ref{eq:conditionalPR1}:
\begin{align}
    \PR(x_1|x_2) = \frac{\denscond{x_1}{x_2,S_1}}{\denscond{x_1}{x_2,S_0}}\,,\label{eq:PR_example}
\end{align}
where the conditional densities are univariate Gaussian  
\begin{align}
    \denscond{x_1}{x_2,S_i} = \gauss{x_1}{ \mu_1(S_i)+\rho\cdot (x_2-\mu_2(S_i))}{ 1-\rho^2 }\,.\label{eq:gausscond}
\end{align}
The mean of the conditional distribution is shifted by $\rho\cdot (x_2-\mu_2(S_i))$ and the variance is reduced by $\rho^2$. Inserting into equation \ref{eq:PR_example}, taking logarithms and re-arranging terms we find
\begin{align}
    \log\PR(x_1|x_2) = \frac{\mu_1(S_1)-\mu_1(S_0)-\rho(\mu_2(S_1)-\mu_2(S_0))}{1-\rho^2}\cdot \left(x_1 - [\overline{\mu}_1+\rho(x_2-\overline{\mu}_2)]\right)\,.\label{eq:logPR_example}
\end{align}
Here, we have denoted the average expectation values by
\begin{align}
    \overline{\mu}_i = ( \mu_i(S_0)+\mu_i(S_1))/2\,.
\end{align}
We choose to discuss the logarithm to slightly simplify the equations. The inflection point $\PR=1$ corresponds to $\log\PR=0$, meaning that the sign of $\log\PR$ determines which scenario is preferred. For $\rho=0$, we recover the unconditional attribution, otherwise different effects can occur depending on the distribution of the conditions.

\subsection{Conditions independent of the scenario}

We can simplify our example further by assuming that the distribution of $X_2$ is the same under both scenarios ($\mu_2(S_0)=\mu_2(S_1)=\mu_2$, figure \ref{fig:toy}\,a,b) -- not an unrealistic setting if $X_2$ represents some aspects of atmospheric dynamics, trends of which tend to be uncertain and small compared to the thermodynamic effects of climate change. From equation \ref{eq:conditionalPR2}, we recognise that the conditional $\PR(x_1|x_2)$ is then equal to the joint $\PR(x_1,x_2)$ since $\PR(x_2)=1$. With this assumption, equation \ref{eq:logPR_example} becomes
\begin{align}
    \log\PR(x_1|x_2) = \underbrace{\left(x_1 - [\overline{\mu}_1+\rho (x_2-\mu_2)]\right)}_{(i)} \cdot \underbrace{(\mu_1(S_1)-\mu_1(S_0))}_{(ii)} \cdot \underbrace{ (1-\rho^2)^{-1} }_{(iii)}\,.\label{eq:logPR_example2}
\end{align}
In this reduced example, the attribution statement depends on three terms:
term $(i)$ measures the deviation of $x_1$ from its mean expectation value $\overline{\mu}_1$, which is shifted up or down according to $x_2$. For example, an unusually high value of $x_1$ may  actually be lower than expected, once we account for the fact that $x_2$ was very large as well. As a result, term $(i)$ can even flip its sign compared to the unconditional case ($\rho=0$).

If the deviation in $(i)$ has the same sign as the difference in expectation between the scenarios $(ii)$, we get $\log\PR>0$, i.e., evidence in favour of the climate change scenario. Conversely, if $(i)$ and $(ii)$ have opposite signs, our belief in $S_0$ would increase. In either case, the attribution statement is strengthened by term $(iii)$ which is always greater than 1 and represents the improvement in signal to noise ratio due to the conditioning on $x_2$. In other words, even conditioning on a perfectly average $x_2=\mu_2$ will increase $\PR$ compared to the unconditional case.  This effect is exemplified in figure \ref{fig:toy}\,(a,b): the separation between the conditional pdfs (curves above the contours) is improved when we increase the correlation from 0.2 in (a) to 0.8 in (b).
In causal inference theory this combination of event (outcome), scenarios (exposure) and storyline condition defines the conditioning variables as moderators (\cite{vanderweele2015}, p.216). 

\subsection{Same climate change signal in event and conditions}
Alternatively, we can consider the case where $X_1$ and $X_2$ experience the same shift in expectation
\begin{align}
    \mu_1(S_1)-\mu_1(S_0) = \mu_2(S_1)-\mu_2(S_0) \,.\label{eq:delta_mu}
\end{align}
For example, we might compute a probability ratio for eastern European temperatures in August 2010 ($X_1$), given that we already know about the preceding hot July ($X_2$). Assuming that the two summer months are equally affected by climate change (equation \ref{eq:delta_mu}), equation \ref{eq:logPR_example} simplifies to
\begin{align}
    \log\PR(x_1|x_2) = \left(x_1 - [\overline{\mu}_1+\rho(x_2-\overline{\mu}_2)]\right)\cdot (\mu_1(S_1)-\mu_1(S_0)) \cdot (1+\rho)^{-1}\,.
\end{align}
The first two terms are the same as in the previous example (equation \ref{eq:logPR_example2}). The effect of correlations (third term), however, is changed: instead of strengthening the attribution statement, a positive correlation reduces $\log\PR$ by a factor of up to two. Intuitively, when the state of $X_2$ is fixed, part of its climate change signal is ``regressed out'' of $X_1$. This effect is shown in figure \ref{fig:toy}\,(c,d).
We would expect such a positive (auto-) correlation in the monthly temperature example above. Here, $\PR$ is further reduced if the preceding month was already hot, but even conditioning on a relatively cool $x_2=\overline{\mu}_2$ would weaken the attribution result compared to $\rho=0$. 
Contrasting the moderator condition from above in causal inference this situation between event (outcome), scenarios (exposure) and storyline condition leads to the characterisation of the condition as mediators (\cite{vanderweele2015},p.216).

\section{Methods}\label{sec:method}
To see the effects of conditional attribution in a real-world example, we will consider the task of attributing summer mean temperature anomalies conditional on atmospheric blocking. This section explains our probability model and blocking definition. 

\subsection{Gaussian mixtures}
Thanks to the efforts of the CMIP community, we have knowledge about the scenarios $S_i$ in the form of multiple climate models $M_{i,j}$, $j=1,...,J$. Each model has its own state vector which we will describe by a random variable $Y$ with realisations $y\in\mathbb{R}^q$. We use a Bayesian approach extending on the work of \cite{min2004bayesian} to combine the information from all models. For conditional and unconditional Bayesian attribution, we need the probabilities of the two scenarios given the observed event $X=x\in\mathbb{R}^q$. This can formally obtained by marginalisation with respect to the model variables $y$ and the models $M_{i,j}, j=1,\ldots,J$:
\begin{align}
    \prob(S_i|x)= \int_{\mathbb{R}^q}\sum_{j=1}^J\mix(S_i,M_{i,j},y|x)~dy\,.\label{eq:margin}
\end{align}
The marginalisation across the models avoids the use of the CMIP ensemble as a single ensemble of opportunity (\cite{tebaldi2007use}).  
The meaning of the joint probability mass and density $\mix$ on the right hand side of this equation becomes clear when we apply Bayes theorem and split up the result into  discrete and continuous components:
\begin{align}
    \mix(S_i,M_{i,j},y|x) &= \frac{\dens(x|S_i,M_{i,j},y)\mix(S_i,M_{i,j},y)}{\dens(x)}\\
    &= \frac{\dens(x|y)\dens(y|S_i,M_{i,j})\prob(S_i)\prob(M_{i,j})}{\dens(x)}\label{eq:hierarchy}
\end{align}
Here we assume that the observations are conditionally independent of the model and scenario given the realisation, i.e., $\dens(x|S_i,M_{i,j},y) = \dens(x|y)$. 

The probabilities $\prob(S_i)$ and $\prob(M_{i,j})$ in equation \ref{eq:hierarchy} represent our a priori belief in each scenario and model and could be chosen at will. The normalisation $\dens(x)$ is independent of the scenario and cancels from the attribution results below when using the probability ration $\PR$.

The term $\dens(x|y)$ represents the distribution of the observations $X$ we would make, given that the climate system is really in the state $y$. We will model this as a multivariate Gaussian density with expectation value $y$ and covariance matrix $\Sigma_o$:
\begin{align}
    \dens(x|y)=\gauss{x}{y}{\Sigma_o}\label{eq:x|y}
\end{align}

The remaining term $\dens(y|S_i,M_{i,j})$ represents the distribution of climate states for climate model $j$ under scenario $i$. Instead of assuming that each model produces a simple Gaussian, we follow the approach of \cite{scholzel2011probabilistic} and employ a mixture of Gaussians, centred at the realisations $y_{i,j,k}$:
\begin{align}
    \dens(y|S_i,M_{i,j}) = \frac{1}{K_{i,j}}\sum_{k=1}^{K_{i,j}}\gauss{y}{y_{i,j,k}}{\Sigma_{i,j}}\label{eq:y|SM}
\end{align}
Inserting equations \ref{eq:hierarchy}, \ref{eq:x|y} and \ref{eq:y|SM} in the original expression (equation \ref{eq:margin}) and rearranging terms, we arrive at
\begin{align}
    \prob(S_i|x) = \frac{\prob(S_i)}{\prob(x)}\cdot \int_{\mathbb{R}^q} \gauss{x}{y}{\Sigma_o} \Big( \sum_{j=1}^J \sum_{k=1}^{K_{i,j}}\frac{\prob(M_{i,j})\gauss{y}{y_{i,j,k}}{\Sigma_{i,j}}}{K_{i,j}} \Big) dy\,.\label{eq:pSo}
\end{align}

The integral can be solved analytically using Woodbury's formula and the matrix determinant lemma, which leads us to
\begin{align}
    \prob(S_i|x) = \frac{\prob(S_i)}{\prob(x)}\sum_{j=1}^J \frac{\prob(M_{i,j})}{K_{i,j}}\sum_{k=1}^{K_{i,j}}\gauss{x}{y_{i,j,k}}{ \Sigma_o+\Sigma_{i,j}}\,.\label{eq:analytic_int}
\end{align}
The probability of each scenario given the the observed event is thus described as a double mixture over multivariate normal distributions centred at the ensemble members, with covariances that add up from the respective models and the observation.

Inserting into equation \ref{eq:PRbayes} yields the probability ratio needed for conditional and unconditional attribution. If  the same climate models are used for both scenarios, and the priors $\prob(M_{i,j})$ are the same for all of them, the end result simplifies to
\begin{align}
    \PR(x) = \frac{\sum_{j,k}(K_{1,j})^{-1}\gauss{x}{y_{1,j,k}}{ \Sigma_o+\Sigma_{1,j}}}{\sum_{j,k}(K_{0,j})^{-1}\gauss{x}{y_{0,j,k}}{ \Sigma_o+\Sigma_{0,j}}}\,.\label{eq:mix_final}
\end{align}

\subsection{Estimation of covariance matrices}
The only remaining difficulty before equation \ref{eq:mix_final} can be applied to data concerns the covariance matrices of the climate models under the two scenarios ($\Sigma_{i,j}$) and the corresponding covariance matrix of the observations $\Sigma_o$. This task is simplified in our bi-variate example case since each matrix is $2\times 2$ with only three independent parameters (the variances and the correlation). If all distributions were represented by sufficiently large ensembles, we could estimate these quantities from the members for each year separately. This is not feasible for most climate models and certainly not for the observations. Instead, we will assume that the covariance values are stationary in time. They could thus in principle be estimated from each timeseries, but that would require us to model and remove the time dependent expectation values first. 

To avoid this additional complication, we further assume that the $\Sigma_{i,j}$ are independent of the scenario and estimate them from the long piControl runs. 
This assumption seems sensible given that we already assume stationarity in time: near the beginning of the timeseries, the historical forcing is almost identical to hist-nat and piControl and the resulting statistics should be interchangeable. If the covariances don't change over time, they should thus remain interchangeable for the whole period. Finally, we assume that all of the $\Sigma_{i,j}$ are equally likely estimates of the ``true'' (observed) variability $\Sigma_o$ and repeat the attribution study for each possible choice. The resulting spread will be reported below.

\subsection{Blocking definition}
For our example application, we want to summarise the presence of atmospheric blocking by a single index which is strongly correlated with surface temperatures, can be localised over the region of interest and does not automatically contain a thermodynamic trend (for example due to the expansion of the atmosphere). These criteria are met by the relatively recent approach of \cite{sousa2021new} based on daily $500\,hPa$ geopotential fields. A key novelty of this blocking definition is the identification of subtropical ridges in addition to classic Rex- or Omega-type blocks. Grid points are considered instantaneously blocked if either (a) the geostrophic flow in $500\,hPa$ is reversed (Rex or Omega, similar to \cite{davini2012bidimensional}) or (b) a ridge with geopotential above the hemispheric mean in the preceding 15 days extends north of the edge of the subtropical belt. Contiguous blocking objects are identified in the daily fields, filtered for sizes $\geq 500.000\,km^2$ and then tracked over time to ensure a lifetime of at least four days. For the details of the algorithm, we refer to \cite{sousa2021new}. The resulting daily binary blocking fields are then averaged over the study region and each summer to obtain a yearly timeseries of blocking activity.

\section{Data}\label{sec:data}

\begin{figure}
    \centering
    \includegraphics[width=8.3cm]{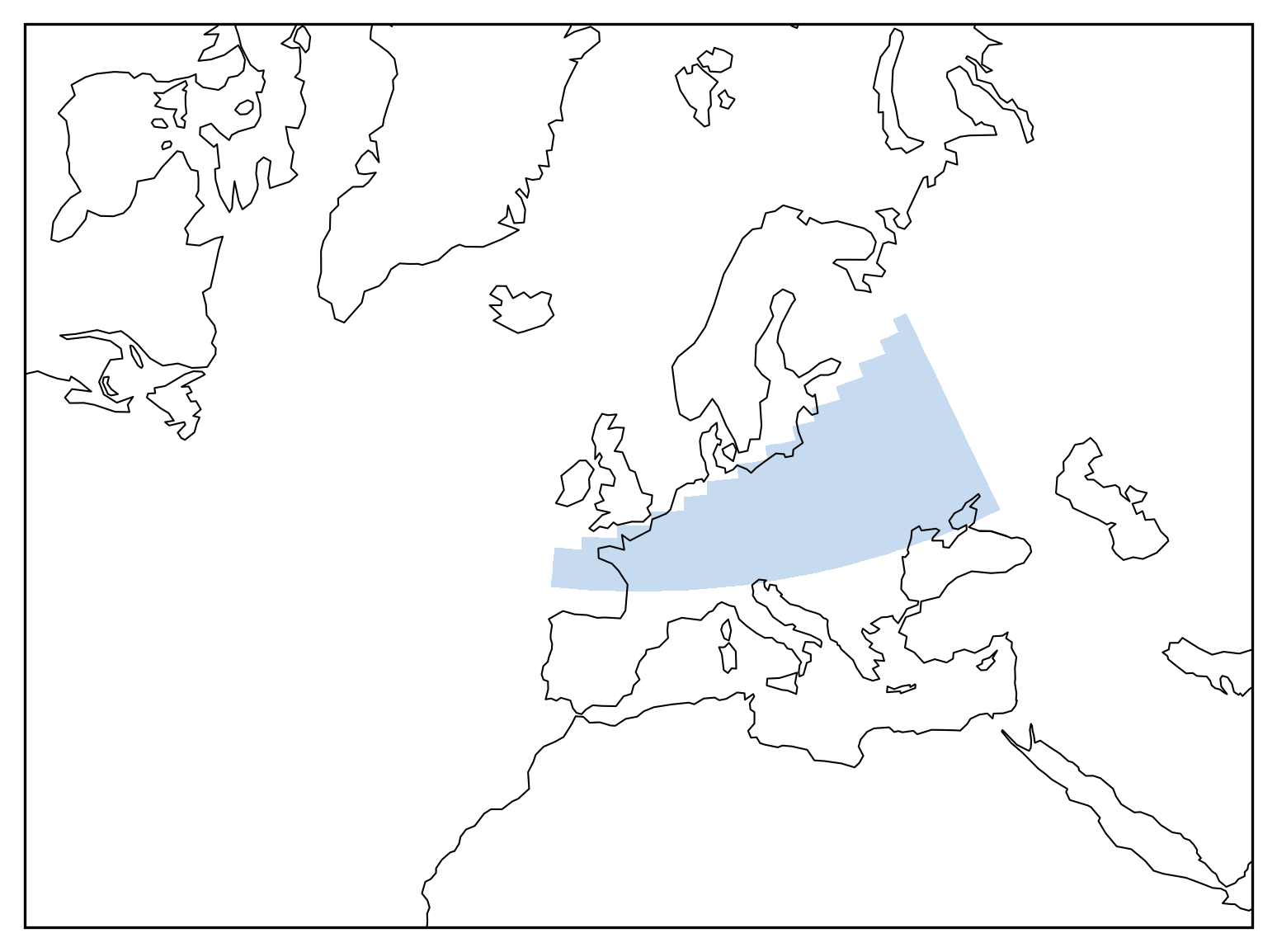}
    \caption{IPCC AR6 central European region CEU \citep{iturbide2020update}.}
    \label{fig:region}
\end{figure}

\begin{table}
    \centering
    \begin{tabular}{lrrr}
        \toprule
        model & hist & hist-nat & piControl \\
        \midrule
        access-cm2 & 3 & 3 & 30 \\
        canesm5 & 35 & 45 & 1051 \\
        cesm2 & 10 & 1 & 1200 \\
        cnrm-cm6-1 & 24 & 3 & 316 \\
        hadgem3-gc31-ll & 4 & 10 & 500 \\
        ipsl-cm6a-lr & 23 & 6 & 115 \\
        miroc6 & 10 & 3 & 500 \\
        mri-esm2-0 & 5 & 5 & 200 \\
        \textit{total} & 114 & 76 & 3912 \\
        \bottomrule
    \end{tabular}
    \caption{Number of hist- and hist-nat members, as well as the number of piControl years available for the CMIP6 ensembles used in this study.}
    \label{tab:data}
\end{table}

We represent the factual and counterfactual scenario by CMIP6 simulations with historical ($S_1$) and hist-nat forcing ($S_0$), respectively. Summer (JJA) mean temperatures are computed from monthly fields and averaged over the IPCC AR6 central European region CEU \citep{iturbide2020update}, shown in figure \ref{fig:region}. Blocking requires us to check the persistence of the flow on daily scales, thus necessitating daily geopotential data from both scenarios as well as piControl. Table \ref{tab:data} shows the eight models for which this data was available. From the two scenarios, we use the time span 1961-2014 which had the largest number of available members. 

Temperatures are converted into anomalies with respect to the reference period 1961-1990, a residual annual cycle is removed from the summer months by subtracting the corresponding reference mean from each month separately. For piControl, the reference means are computed over the entire timeseries.

To ensure physical consistency between surface temperatures and upper level flow, we use ERA5 reanalysis data \citep{hersbach2020era5} to represent the observed state. The ERA5 data is processed in the same way as the historical CMIP6 simulations. 

\begin{figure}
    \centering
    \includegraphics[width=8.3cm]{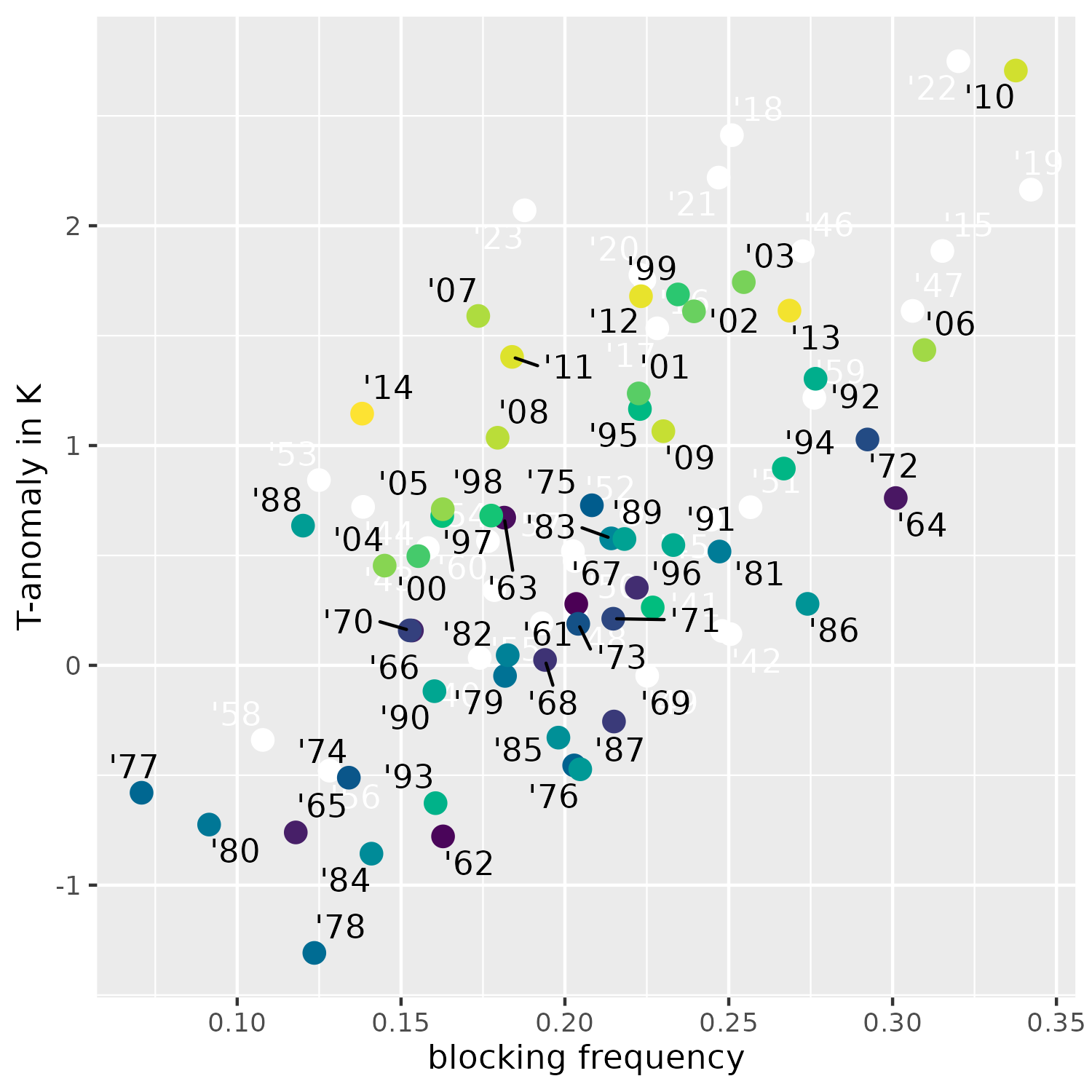}
    \caption{ERA5 JJA mean blocking intensity and temperature anomaly in the CEU region (figure \ref{fig:region}). Colours correspond to the year, white dots in the background are outside the range of the used CMIP6 data and is not attributed in this study. }
    \label{fig:scatter}
\end{figure}

Figure \ref{fig:scatter} summarises the observed temperature anomalies and blocking activity. Besides the expected increase in temperatures over time, we find a strong linear connection to the chosen blocking index. For example, it is interesting to note that the two hottest years before 1999 (1946 and 1947) both exhibit exceptional blocking.  No significant (linear) trend was found in the blocking time series ($p>10\,\%$). Together, linear trend and blocking index explain roughly $70\,\%$ of the temperature variance in the period 1961-2014. 

\begin{figure*}
    \centering
    \includegraphics[width=\textwidth]{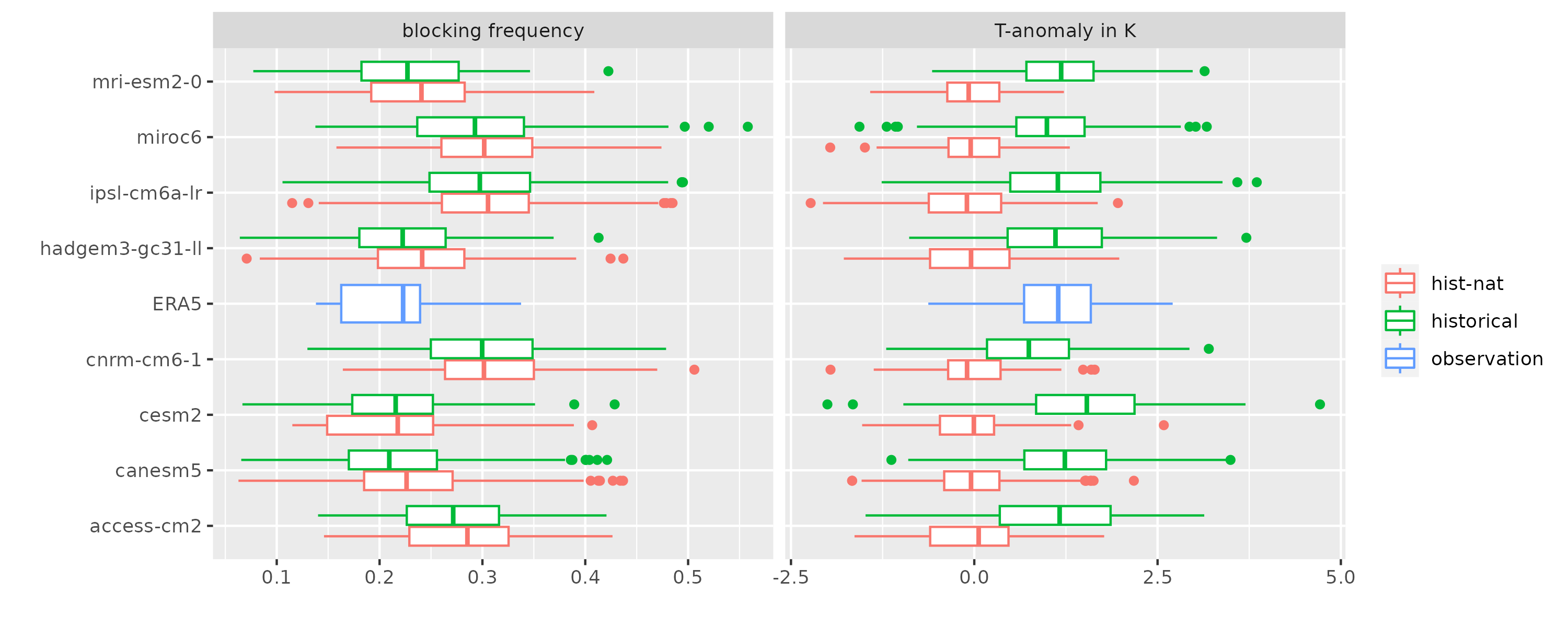}
    \caption{Distribution of blocking fractions  (left) and Temperature anomalies  (right) for the scenario simulations and ERA5 during the period 1990-2014.}
    \label{fig:data_dists}
\end{figure*}

The distributions of temperature and blocking in the CMIP6 data are summarised in figure \ref{fig:data_dists}. Here, we focus on the period 1990-2014, where a clear difference between the scenarios has emerged. Consequently, temperatures in the historical simulations and ERA5 are around $1~K$ warmer than the reference period; the hist-nat runs show no significant anomalies. The agreement between models and ERA5 is substantially worse for the blocking frequency (right panel), where many of the simulations show significantly higher values than ERA5. Differences between the two scenarios are minor, with only some models showing a slight reduction in blocking for the historical runs. 
In principle, information on these biases with respect to ERA5 could be included in the prior probabilities $\prob(M_{i,j})$ in equation \ref{eq:analytic_int}, to down-weight or exclude less reliable simulations. Considering the already limited number of available models, we choose to   keep all models included in the present analysis.

\section{Attribution of summer temperatures conditional on blocking}\label{sec:res}

Conditional and unconditional attributions are carried out for each year in the time series separately. To gauge the uncertainties resulting from our selection of models and the unknown choice of $\Sigma_o$, the entire experiment is repeated 800 times: for each of the eight choices of $\Sigma_o=\Sigma_i$, we additionally draw 100 samples with replacement among the eight ensembles used to represent the scenarios. 

\begin{figure*}
    \centering
    \includegraphics[width=\textwidth]{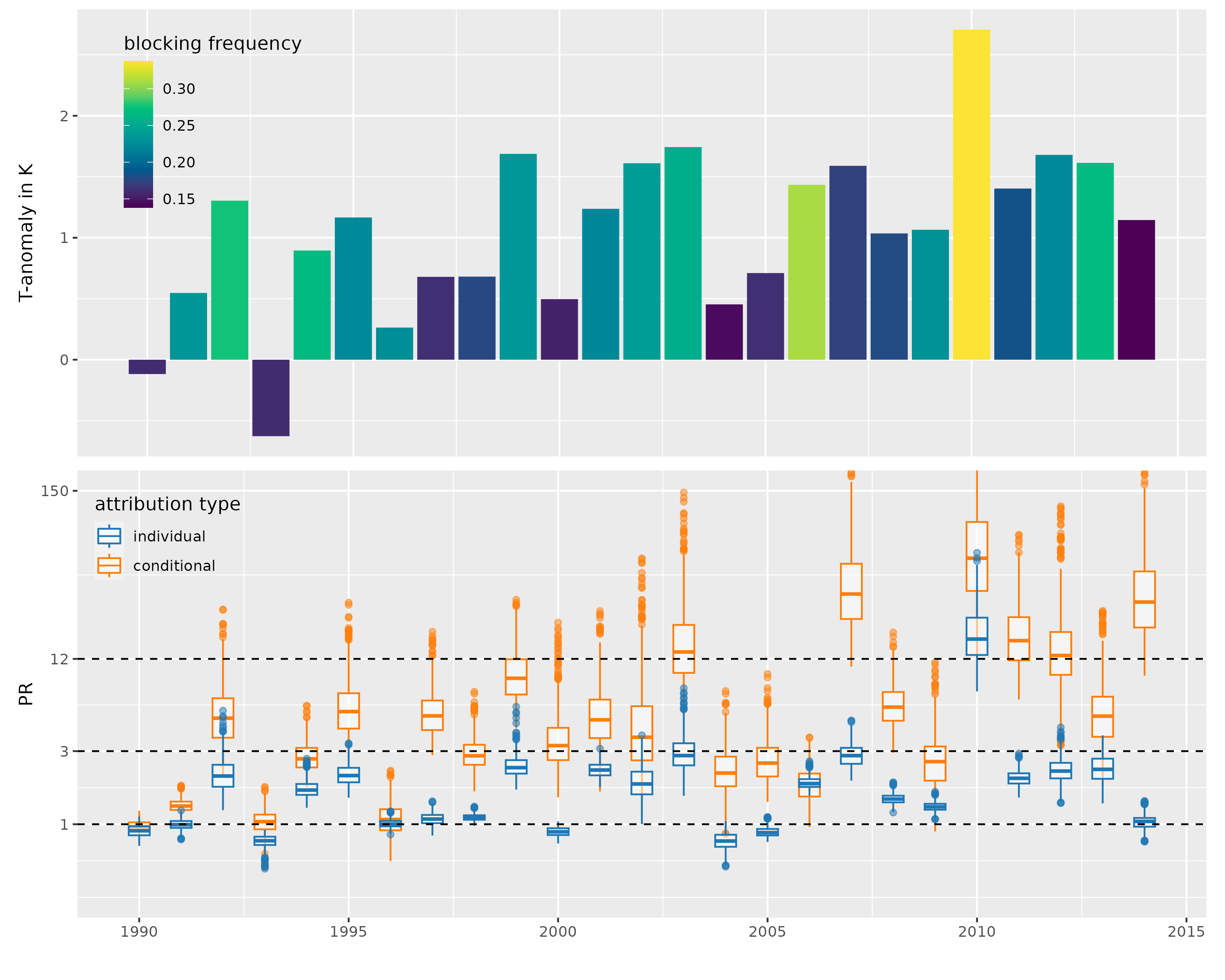}
    \caption{ERA5 temperature anomalies and blocking (top, as in figure \ref{fig:scatter}) and probability ratios for temperature individually and conditional on blocking (bottom, logarithmic y-scale). Box-plots represent sampling uncertainty over the climate models and uncertainty of the observed covariance matrix (see text).}
    \label{fig:PRdist}
\end{figure*}

Figure \ref{fig:PRdist} summarises the attribution results for all years after 1990. Before the 1990s, the two scenarios are very similar and no probability ratios substantially above 1 were found. The box-plots represent the aforementioned sampling uncertainties.

In terms of unconditional attribution (blue boxes), only the summers of 2003, 2007 and 2010 reach or exceed the threshold of $\PR=3$ indicating substantial evidence against the hist-nat scenario; 2010 is the only year for which a ``strong'' level of evidence is found. The summer of 1999 has a comparable temperature anomaly to 2003 but attribution is not achieved because the ensembles are not as well separated yet. The sampling uncertainties are reasonably small compared to the overall range of probability ratios.

Conditioning on the blocking activity almost universally increases the probability ratio, indicating stronger attribution statements. In this setting more than half of the years after 1990 yield at least substantial  evidence, the first one being 1992. In addition to 2010, the summers of 2007 and 2014 are clearly in the ``strong'' category due to their weak blocking activity. 2014 in particular shows a relatively modest temperature anomaly just over $1\,K$. Once we control for the very weak contribution from blocking, this anomaly is revealed to be highly unlikely under $S_0$.  

\begin{figure}
    \centering
    \includegraphics[width=8.3cm]{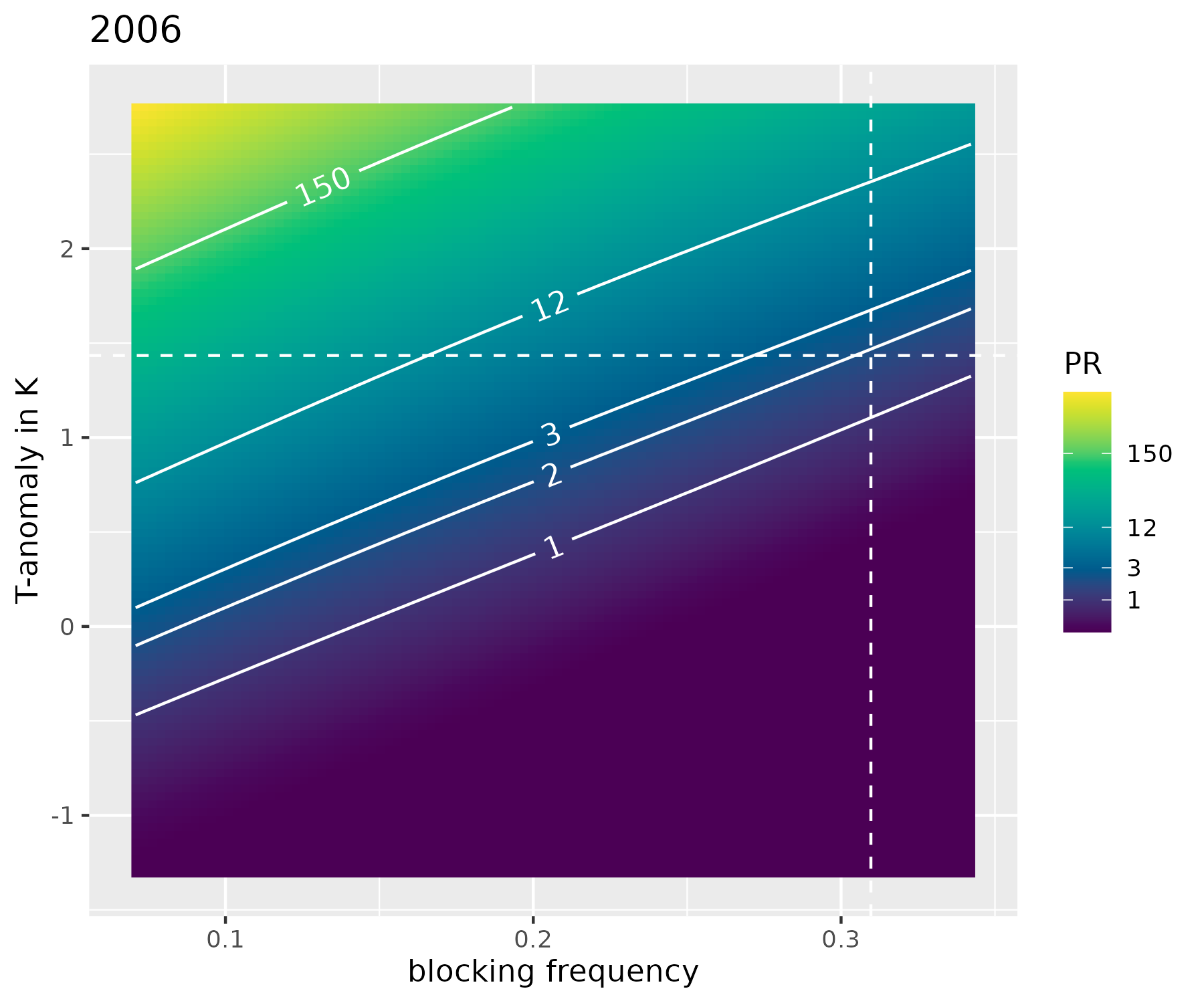}
    \caption{Conditional probability ratio in summer 2006 as a function of observed blocking activity and temperature anomaly (logarithmic colour scale). Dashed lines indicate the values that actually materialised in 2006. To obtain a single realisation of $\PR$, $\Sigma_o$ was set to the mean over all $\Sigma_i$ for this figure.}
    \label{fig:PR06}
\end{figure}

The summer of 2006 is one of the few examples where $\PR$ is not enhanced by the conditioning. Here, the strong positive blocking anomaly explains much of the observed temperature anomaly and $\PR$ is reduced. Figure \ref{fig:PR06} visualises the situation. In this example, even a slightly above average blocking value of 0.25 would have led to substantial evidence for $S_1$ over $S_0$. Conversely, for very weak blocking $<0.1$, even a temperature anomaly near zero would have led to $\PR>3$.  

The fact that spread in $\PR$ (size of the box-plots in figure \ref{fig:PRdist}) is generally increased compared to the unconditional approach is hardly surprising: on the one hand, higher dimensions leave more room for uncertainties. On the other hand, we have seen in figure \ref{fig:data_dists} that the agreement between models is better for temperatures than blocking. 

Lastly, we may wonder whether the increase in $\PR$ is strongly impacted by the positive blocking biases seen in figure \ref{fig:data_dists} for some of the models: based on equation \ref{eq:logPR_example2} we would indeed expect that an increased $\mu_2$ (expectation value of the conditions under both scenarios) would raise $\PR$ by making the observed temperature anomaly seem even more unlikely. We have tested and excluded this possibility by repeating the analysis without acces-cm2, cnrm-cm6-1, ipsl-cm6a-lr and miroc6 (the four models with strong blocking biases in figure \ref{fig:data_dists}). The resulting conditional probability ratios (not shown) are only slightly reduced compared to figure \ref{fig:PRdist}, leaving our conclusions unchanged. 

\section{Connection to storylines and analogues}\label{sec:story}
Before we discuss these results and draw our conclusions, it is worth clarifying how exactly our attribution study is connected to other styles of conditional attribution found in the literature.

Many attribution studies following the storyline approach consider a special case of conditional attribution, wherein the conditions $C$ are experimentally forced to approach some predetermined values, for example when a regional model is set up to simulate a specific event. This can be realised via data assimilation (for example \cite{vanGarderen2020}) or purely via the initial and boundary conditions. In the notation of causal theory \citep{pearl2009causality}, probabilities with experimentally fixed conditions are denoted by $\prob(E|\textit{do}(C))$, which is generally not identical to $\prob(E|C)$. The probability ratio then has the form
\begin{align}
    \PR(E|\textit{do}(C)) = \frac{\prob(E|\textit{do}(C),S_1)}{\prob(E|\textit{do}(C),S_0)}\,.\label{eq:conditionalPRdo}
\end{align}
Equation \ref{eq:conditionalPRdo}  yields the same result as the original equation  \ref{eq:conditionalPR1} if $X_2$ has no causal factors in common with $X_1$ (being a moderator). In particular, this implies that $X_2$ must be independent of the scenario. Otherwise these are two slightly different versions of the attribution question:
\begin{enumerate}
    \item How much more likely is $E$ in scenario $S_1$, given that $C$ is also observed? (equation \ref{eq:conditionalPR1})
    \item How much more likely is $E$ in scenario $S_1$, if $X_2$ is experimentally fixed to be within some range of observed conditions $\setsymbol{C}$? (equation \ref{eq:conditionalPRdo})
\end{enumerate}
In this study, we have asked the former question and used equation \ref{eq:conditionalPR2} to answer it. It should be noted that a storyline analysis may focus on the mean state instead of the full conditional distribution. Our conclusions about stronger attribution statement due to better signal to noise ratio still applies, if we think about the probability that any particular storyline study will find a significant result. 

While the focus of this study is on reconciling the two main streams of attributions studies, another active line of research has recently explored attribution based on analogue methods (see for example \cite{yiou2007inconsistency}, \cite{faranda2022climate}). For a given event of interest, this approach searches for days with analogous weather patterns, for example based on the sea level pressure maps. The climate for analogues from the distant past (representing $S_0$) is then compared to that for analogues from a more recent period ($S_1$). Depending on the number of analogues selected, one could compute the probability of some threshold exceedance or study only the difference climatological mean. 

From the point of view taken in this paper, it is clear that the analogue-method performs a conditional attribution in the sense of equation \ref{eq:conditionalPR1}, i.e., the conditions are observed and not experimentally enforced. 
The condition set $\setsymbol{C}$ of the analogue approach consists of all those states, for which the circulation was sufficiently similar to that of the event (see also \cite{yiou2017statistical}). Since  close analogues in the atmosphere are rare \citep{lorenz1969atmospheric}, this approach leaves considerable room for different states of the overall system.  

In contrast, the condition set $\setsymbol{C}$ for a typical storyline study contains only those system configurations that can be reached by the climate model given the prescribed initial and/or boundary conditions of the event. In  this sense, storyline studies tend to have the narrowest set of conditions (trajectories started near a single point in phase space), followed by analogues (similarity to a specific circulation state) and the conditional approach of this study (some aggregated property of the circulation has a specific value). At the other end of the spectrum is the classic risk-based attribution \citep{philip2020protocol} which sometimes conditions on a large-scale climate index like the global mean temperature anomaly or ENSO or uses no conditions apart from the scenario.

\section{Discussion and conclusion}\label{sec:outro}

The classification of different styles of attribution according to the degree of conditioning is not new \citep{shepherd2016}. In this study, however, we have taken the idea literally and shown how conditional and unconditional attribution are related.  The core result is given by equation \ref{eq:conditionalPR2}, which has enabled us to study the effects of conditioning on paper in a simple toy model (section \ref{sec:toy}) before confirming the results in a more realistic setting. 

We have learned that the attribution statement is strengthened when we choose conditions that are independent of the scenario (moderators) and explain a large part of the variance in the event variable. This effect was strongly pronounced in our real-world example of temperature attribution conditional on blocking: on a seasonal scale, the blocking index of \cite{sousa2021new} explains much of the European summer temperature variance without showing any significant trend itself. When the weak blocking conditions near the end of the timeseries are thus taken into account, even June-July-August 2014 data that were hardly remarkable in terms of temperature alone show clear indications of a changing climate. 

This strengthening of $\PR$ is analogous to the goal of storyline studies: arrive at a less ambiguous attribution statement by removing the uncertainty inherent in the (rare) dynamical conditions. In addition, an analysis that explicitly includes the weather conditions is more directly linked to the specific observed event than a risk-based study considering only a threshold exceedance. 

Analogous considerations apply to the analogue style of attribution, which is even more closely linked to our Bayesian approach: in both cases, the conditional statistics are estimated from observed or modelled time series. Indeed,  the terms in equation \ref{eq:conditionalPR1} could just as well be estimated from a sufficiently large database of analogues. As mentioned above, storylines ask a slightly different attribution question by enforcing the conditions experimentally (equation \ref{eq:conditionalPRdo}), the $\textit{do}(C))$ operation according to \cite{pearl2009causality}.  

Both analogues and storylines can also experience the inverse effect, i.e., a weaker attribution statement when the conditions already contain part of the climate change signal. For the interpretation of these conditional attribution studies, possible past or future trends in the conditions are thus of great importance. In the present study, $C$ is given by a single index, which can easily be tested for trends. Analogue studies can achieve the same by tracking the quality of analogues over time. The situation is more complex for storylines, where $C$ contains some very limited region in the full state space of the atmosphere. Here, one would first need to assess which aspects of the conditions were actually relevant to the event and then investigate their climate change signal in detail.  

Lastly, while the general result given by equation \ref{eq:conditionalPR2} is not limited to a specific statistical model, it is worth mentioning that our fully Bayesian approach  has several attractive properties: the fact that we already know about climate change as a whole is conceptually reflected in the prior probabilities (equation \ref{eq:PRbayes}). Instead of starting with a null hypothesis of ``no climate change'', an individual attribution study merely adds to (or subtracts from) this overall knowledge. 
Furthermore, the Bayesian mixture model allows us combine information from multiple climate model ensembles in a well-defined way. If we have some prior knowledge about the credibility of individual models, we can include it in the prior probabilities $\prob(M_{i,j})$ in equation \ref{eq:analytic_int}. Exclusion of models that are known to be inadequate (as advised by \cite{philip2020protocol}) corresponds to $\prob(M_{i,j})=0$, but intermediate ratings are also be possible. 

In future studies, our model can be extended to higher dimensions with relative ease: as long as the assumption of Gaussian distributions for individual ensemble members is reasonable, the only (although not inconsiderable) difficulty lies in the estimation of the covariance matrices. In this way, compound events in space and time could be attributed, conditional on one or more aspects of internal climate variability.
A next step would consist of moving beyond normal distributions, potentially enabling us to attribute extremes on smaller spatio-temporal scales. The end result may then no longer be analytically tractable (as in equation \ref{eq:analytic_int}), but could be obtained numerically.

\appendix

\section{Probability ratio for thresholds and individual values}\label{app:PRcont}
In section \ref{sub:conti} we clarified how probability ratios can be computed for individual continuous values $X=x$. It is a natural question whether the result of such an attribution will be stronger or weaker compared to the usual definition $X>x$ \citep{philip2020protocol}. We can write the probability of a threshold exceedance (also called survival function) by $\prob(X>x) = 1-F(x)$, where $F(x)$ is the cumulative distribution function, and compare the two probability ratios
\begin{align}
    \frac{\PR(X>x)}{\PR(X=x)}= \frac{ 1-F(x|S_1) }{ \denscond{x}{S_1} } ~ \frac{\denscond{x}{S_0}  }{ 1-F(x|S_0) } = \frac{ \frac{\partial}{\partial x}\ln\left( 1-F(x|S_0) \right) } {\frac{\partial}{\partial x}\ln\left( 1-F(x|S_1) \right)}\,,\label{eq:PRlogtail}
\end{align}
where we have used the fact that $\dens$ is the derivative of $F$ in the last step. Thus, the change in $\PR$ when we switch from a threshold-based event to a continuous event depends on the slope of the logarithmic survival function.


Attribution studies for extreme events often rely on the generalised extreme value (GEV) family of distributions. To simplify the notation, we introduce the re-scaled variable
\begin{align}
    \tau(x) = \frac{x-m}{s}\,,
\end{align}
where $m$ and $s$ are the location and scale parameter respectively. 
In the case of a Gumbel distribution (GEV with neither upper nor lower endpoint), we then have for large $x$
\begin{align}
    \ln\left( 1 - F_\textit{Gu}(x|S_i) \right) = \ln\left( 1 - e^{-e^{-\tau(x))}} \right)\approx \ln\left( e^{-\tau(x)} \right) = -\tau(x)\,,
\end{align}
where we have used the Taylor-Series $1-e^{-y}\approx y$ for small $y$ (in our case large $x$). Inserted into equation \ref{eq:PRlogtail}, we obtain in the Gumbel case for large $x$
\begin{align}
    \frac{\PR(X>x)}{\PR(X=x)}\approx \frac{s(S_1)}{s(S_0)}\,.\label{eq:PrGu}
\end{align}
For the Gumbel distribution, this can also be expressed as the ratio between the standard deviations $\sigma(S_i)$ since $s$ is proportional to $\sigma$.

For the Fr\'echet case (heavy tailed GEV), we can use the same Taylor series to find
\begin{align}
    \ln\left( 1 - F_\textit{Fr}(x|S_i) \right) = \ln\left( 1 - e^{-\tau(x)^{-\alpha}} \right)\approx \ln\left( \tau(x)^{-\alpha} \right) = \alpha\ln(\tau(x))
\end{align}
and thus
\begin{align}
    \frac{\PR(X>x)}{\PR(X=x)}\approx \frac{\alpha(S_0)}{\alpha(S_1)}\cdot \frac{x-m(S_1)}{x-m(S_0)}\,.\label{eq:PRFr}
\end{align}
In the limit of large $x$, the second factor approaches unity and the ratio between the two attribution results is again a constant factor, this time depending only on the shape parameter $\alpha$. 

The event attribution protocol of \cite{philip2020protocol} typically models temperature extremes as a GEV distribution with constant shape and scale, where only the location parameter $m$ depends on the scenario (parametric function of the global mean surface temperature anomaly). According to equations \ref{eq:PrGu} and \ref{eq:PRFr}, the attribution results for thresholds and individual values are then identical in the limit of large $x$, as long as a Gumbel or Fr\'echet distribution are assumed (in the Weibull case, such a limit makes no sense because of the upper endpoints). In the Fr\'echet case, this remains true even when the scale parameter $s$ is variable, as is sometimes done for the attribution of heavy precipitation events \citep{philip2020protocol}.

\section*{Author contributions}
AH and PF obtained the funding and supervised this work. AH derived the relationship between storylines and risk-based attribution and developed the Gaussian mixture model. SB refined the mathematics and led the data handling, computation,  visualisation and writing. All authors contributed to the review and editing of the paper.

\section*{Acknowledgements}
This research was funded within the BMBF project ClimXtreme phase 1 under grant number 01LP1902A and  ClimXtreme phase 2 under grant number 01LP2323A. We  gratefully acknowledge DKRZ for the computational resources and data access granted and especially to Etor Lucio Eceiza for his invaluable technical support. We furthermore thank Ieda Pscheidt-Willems for her contributions in the early phases of this work. 

\bibliography{sources}

\end{document}